\documentclass[twocolumn,aps,showpacs,prb,tightenlines,amsmath,amssymb,superscriptaddress]{revtex4}
\usepackage{graphicx}

\usepackage{dcolumn}

\usepackage{bm}

\usepackage{colordvi}

\begin{document}

\title{Dependence of spin dephasing on initial spin polarization in
a high-mobility two-dimensional electron system}

\author{D.\ Stich}
\affiliation{Institut f\"ur Experimentelle und Angewandte Physik,
Universit\"at Regensburg, D-93040 Regensburg, Germany}
\author{J.\ Zhou}
\affiliation{Hefei National Laboratory for Physical Sciences at
Microscale and Department of Physics, University of Science and
Technology of China, Hefei, Anhui, 230026, China}
\author{T.\ Korn}
\affiliation{Institut f\"ur Experimentelle und Angewandte Physik,
Universit\"at Regensburg, D-93040 Regensburg, Germany}
\author{R.\ Schulz}
\affiliation{Institut f\"ur Experimentelle und Angewandte Physik,
Universit\"at Regensburg, D-93040 Regensburg, Germany}
\author{D.\ Schuh}
\affiliation{Institut f\"ur Experimentelle und Angewandte Physik,
Universit\"at Regensburg, D-93040 Regensburg, Germany}
\author{W.\ Wegscheider}
\affiliation{Institut f\"ur Experimentelle und Angewandte Physik,
Universit\"at Regensburg, D-93040 Regensburg, Germany}
\author{M.\ W.\ Wu}
\email{mwwu@ustc.edu.cn.} \affiliation{Hefei National Laboratory for
Physical Sciences at Microscale and Department of Physics,
University of Science and Technology of China, Hefei, Anhui, 230026,
China}
\author{C.\ Sch\"uller}
\email{christian.schueller@physik.uni-regensburg.de.}
\affiliation{Institut f\"ur Experimentelle und Angewandte Physik,
Universit\"at Regensburg, D-93040 Regensburg, Germany}

\date{\today}

\begin{abstract}
We have studied the spin dynamics of a high-mobility two-dimensional
electron system in a GaAs/Al$_{0.3}$Ga$_{0.7}$As single quantum well
by time-resolved Faraday rotation  and time-resolved Kerr rotation
in dependence on the initial degree of spin polarization, $P$, of
the electrons. By increasing the initial spin polarization from the
low-$P$ regime to a significant $P$ of several percent, we find that
the spin dephasing time, $T_2^\ast$, increases from about 20\ ps to
200\ ps; Moreover, $T_2^\ast$ increases with temperature at small
spin polarization but decreases with temperature at large spin
polarization. All these features are in good agreement with
theoretical predictions by Weng and Wu [Phys. Rev. B {\bf 68},
075312 (2003)]. Measurements as a function of spin polarization at
fixed electron density  are
performed to further confirm the theory. A fully microscopic
calculation is performed by setting up and
 numerically solving the kinetic spin Bloch equations, including the D'yakonov-Perel' and the
Bir-Aronov-Pikus mechanisms, with {\em all}
 the scattering explicitly included. We reproduce all principal features of
the experiments,
 i.e., a dramatic decrease of spin dephasing with increasing
$P$ and the temperature dependences at different
spin polarizations.
\end{abstract}


\maketitle
\section{Introduction}
In the past decade, the spin degrees of freedom in semiconductors
have been investigated both experimentally and theoretically due to
the great prospect of potential applications in spintronics or
quantum computational devices. \cite{Awschalom1,Fabian}
Unaffectedly, the study of spin dephasing/relaxation has been one of
the most important and interesting branches of this field. The
dominant spin dephasing mechanism in $n$-doped GaAs quantum wells
(QWs) is the D'yakonov-Perel' (DP) mechanism.\cite{dp} It is caused
by the $k$-vector dependent effective magnetic fields which arise
from the bulk inversion
 asymmetry (BIA)\cite{bia} and
the structure inversion asymmetry (SIA).\cite{sia}  Electrons with different
$k$-vectors experience  different effective magnetic fields
$\textbf{B}(k)$, and they
 would precess at different frequency $\Omega$. By averaging the
magnitude of $\textbf{B}(k)$ over the momentum distribution of the
electrons, an average Larmor frequency $\Omega_{av}$ due to the
effective magnetic fields can be determined. For the DP mechanism,
two limiting cases are considered:\cite{Fabian,lvchengwu} (i) $\tau_p
\Omega_{av} \geq 1$: If the product of average Larmor frequency
$\Omega_{av}$ and momentum relaxation time $\tau_p$ is larger than
one, spins may precess more than a full cycle
before being scattered into another momentum state.
Strong interference induced decay happens in this limit.
(ii) $\tau_p \Omega_{av} \leq 1$: in this
regime, the momentum relaxation time $\tau_p$ is so short that the
effective magnetic field $\textbf{B}(k)$ may be treated as a rapid
fluctuation. Individual electron spins only precess by a fraction of
a full cycle before the effective magnetic field changes amplitude
and direction due to momentum scattering.
In this regime, the spin dephasing time $\tau_s$ is inversely
proportional to the momentum relaxation time $\tau_p$. This behavior
is commonly called motional narrowing.

In a number of experiments performed by Kikkawa {\em et
al.},\cite{Kikkawa1,Kikkawa2} extremely long spin relaxation times
could be achieved in GaAs bulk material\cite{Kikkawa1} or in II-VI
quantum wells\cite{Kikkawa2} by using doping levels close to the
metal-insulator transition. The spin dephasing close to the
metal-insulator transition was further studied by Sandhu {\em et
al.}.\cite{sandhu} The dopants act as centers of momentum scattering
that enhance spin lifetime due to motional narrowing. On one hand,
this is helpful for manipulation of optically-excited spins. On the
other hand, however, a high impurity density is undesirable for a
transistor device, where highly-mobile charge carriers are required
with dissipation processes as low as possible. Most proposals for
spin transistor device structures are within the ballistic regime
and thus require extremely high mobility. Recently, however,
Schliemann {\em et al}. introduced a concept for a spin transistor
device working in the diffusive transport regime.\cite{Schliemann03}
Measurements of the spin dephasing in modulation n-doped quantum
wells have so far focused on structures grown in the $[110]$ crystal
direction, in which Ohno {\em et al}. found spin dephasing times of
several nanoseconds.\cite{Ohno} These are due to the fact that in a
$[110]$-grown QW, the  Dresselhaus spin-orbit field points along the
growth direction for electrons of arbitrary $k$ vector. For electron
spins aligned along the $[110]$ direction, the DP mechanism is thus
absent, as the spins are parallel to the Dresselhaus field and do
not precess. As soon as a magnetic field is applied in the sample
plane, however, the spins are forced to precess and change their
orientation. This leads to a drastic decrease of the spin dephasing
time (SDT), shown theoretically by Wu and Kuwata-Gonokami\cite{wugo}
and experimentally by D\"ohrmann {\em et al.}.\cite{Dohrmann} This
is because for spins with an orientation different from $[110]$, the
Dresselhaus field again causes a precessional motion, leading to
dephasing due to the DP mechanism. From the point of view of
applications, the  advantage of long spin dephasing time in
$[110]$-grown QWs is thus diminished, as the manipulation of spins
by an external magnetic field destroys it. This is also the case for
electrical fields applied in the growth direction, either by
asymmetric modulation doping of the QW, or by an external gate
voltage. This is due to the fact that an electrical field induces a
structural inversion asymmetry, which manifests itself in the Rashba
spin-orbit term in the Hamiltonian. Like the Dresselhaus term, it
may be described by a $k$-dependent effective magnetic field. For
the $[110]$-grown QW, the Rashba field direction is within the
sample plane, thus causing  spin dephasing even for spins aligned in
the $[110]$ direction. Karimov {\em et al}. demonstrated that the
SDT in a $[110]$-grown QW may be decreased by an order of magnitude
by applying an electrical field in the growth direction which
effectively tunes the contribution of the Rashba field to spin
dephasing.\cite{Karimov}

In high-mobility two-dimensional electron systems (2DES),
electron-electron Coulomb interaction can play an important role. It
was first pointed out by Wu and Ning\cite{Wu} that any scattering
including the spin conserving Coulomb scattering can cause an
irreversible spin dephasing in the presence of inhomogeneous
broadening. This inhomogeneous broadening can be from the
energy-dependent $g$-factor,\cite{Wu} the Dresselhaus/Rashba
terms,\cite{wu3d} and even the $k$-dependent spin diffusion along a
spacial gradient.\cite{wutr}
 Recently, also for $[001]$-grown $n$-doped QWs, the importance of the electron-electron scattering for spin
relaxation and dephasing was proved by Glazov and Ivchenko\cite{Ivchenko}
by using perturbation theory and by Weng and Wu\cite{wu1} from a fully
microscopic many-body approach. In a
thorough temperature-dependence study of the spin dephasing in
$[001]$-oriented $n$-doped QWs, Leyland {\em et al}.
 experimentally verified the effects of the
electron-electron scattering.\cite{Leyland07} In almost all
theoretical and experimental investigations, the spin polarization
is very small and there is no/small external electric field parallel
to the QWs. In other words, the spin systems are near the
equilibrium. Nevertheless, Wu {\em et al.} set up the kinetic spin
Bloch equations which can be used to investigate the spin kinetics
regardless of how far away from the
equilibrium.\cite{wumetiu,wurev,wu1,wu2} While numerically solving
these equations, all the scatterings such as electron-acoustic
phonon, electron-longitudinal phonon, electron-nonmagnetic impurity,
and especially the electron-electron Coulomb scatterings are
explicitly included.\cite{wu1,wu2,wu3} Weng and Wu predicted an
interesting effect that the spin dephasing is greatly suppressed by
increasing the initial spin polarization in Ref.\ \onlinecite{wu1}.
This effect comes from the Hartree-Fock (HF) term of the Coulomb
interaction. This term serves as an effective magnetic field along
the $z$-axis which can be greatly increased with the spin
polarization and therefore blocks the spin precession as a result of
the lack of detuning.\cite{wu1} Moreover, they further predicted
that for high mobility samples, the spin dephasing time decreases
with temperature at high spin polarization, which is in {\em
opposite} to the case of small polarization.\cite{wu1}

Here, we report on time-resolved experiments in which we manage to
realize a significant spin polarization and indeed observe the
proposed effects. Spin-polarized carriers are injected into the 2DES
at the Fermi level by way of optical pumping with a
circularly-polarized laser. The SDT $T_2^{\ast}$ is determined
through time-resolved Faraday rotation (TRFR) and time-resolved Kerr
rotation (TRKR). We find that $T_2^{\ast}$ visibly increases with
increasing initial spin polarization and it increases/decreases with
temperature at small/large spin polarization. All these features are
in good agreement with the theoretical predictions.\cite{wu1} In
addition, we present the effect of spin-conserving and spin-flip
electron-heavy hole scattering, and the screening from the hole gas
on spin dephasing.  Control experiments using constant
excitation density and varying the circular polarization degree of
the pump beam demonstrate that the observations are due to an
increased initial spin polarization instead of  caused by either
increased electron density or changes in sample temperature.
Moreover, the variation of the electron $g$ factor with degree of
spin polarization has the same tendency both in experiment and
theory.

This paper is organized as following. In Sec.\ II, we construct the
kinetic spin Bloch equations. Then we describe the preparation of
the sample in Sec.\ III. The setup of the experiment and the main
results both in experiments and calculations are presented in Sec.\
IV. We conclude in Sec.\  V.

\section{Microscopic calculations}

First, we construct the kinetic spin Bloch equations in GaAs QWs by
using the nonequilibrium Green's function method:\cite{haug}
\begin{equation}
\dot{\rho}_{{\bf k},\sigma \sigma^{\prime}}=\dot{\rho}_{{\bf k},
\sigma \sigma^{\prime}}|_{\mbox{coh}}
+\dot{\rho}_{{\bf k},\sigma \sigma^{\prime}}|_{\mbox{scatt}}\ ,
\label{bloch}
\end{equation}
with $\rho_{{\bf k},\sigma \sigma^{\prime}}$ representing the single
particle density matrix elements. The diagonal and off-diagonal
elements of $\rho_{{\bf k},\sigma \sigma^{\prime}}$ give the
electron distribution functions $f_{{\bf k}\sigma}$ and the spin
coherence $\rho_{{\bf k},\sigma-\sigma}$, respectively. The coherent
terms $\dot{\rho}_{k,\sigma \sigma^{\prime}}|_{\mbox{coh}}$ describe
the precession of the electron spin due to the effective magnetic
field from the Dresselhaus term\cite{dress} $\mathbf{\Omega}^{\mbox
{\small BIA}}({\bf k})$, the Rashba term $\mathbf{\Omega}^{\mbox
{\small SIA}}({\bf k})$, and the HF term of Coulomb interaction. The
expressions of the coherent term can be found in Refs.\
\onlinecite{wu1,wu2}. The Dresselhaus term can be written
as:\cite{dp1} $ \Omega^{\mbox {\small BIA}}_x({\bf k})=\gamma
k_x(k_y^2-\langle k_z^2\rangle) $, $\Omega^{\mbox {\small
BIA}}_y({\bf k})=\gamma k_y(\langle k_z^2\rangle-k_x^2)$, and,
$\Omega^{\mbox {\small BIA}}_z({\bf k})=0$, in which $\langle
k_z^2\rangle$ represents the average of the operator
$-(\partial/\partial z)^2$ over the electronic state of the lowest
subband.\cite{wu3} $\gamma$ is the spin splitting
parameter,\cite{meier} and we choose it to be $17.1$\
eV$\cdot$\AA$^3$ all through the paper. The Rashba term can be
written as: $\Omega^{\mbox {\small SIA}}_x({\bf k})=\alpha k_y$,
 $\Omega^{\mbox {\small SIA}}_y({\bf k})=-\alpha k_x$, and,
$\Omega^{\mbox {\small SIA}}_z({\bf k})=0$, in which the Rashba
spin-orbit parameter $\alpha$ is proportional to the interface
electric field, and we choose it to be $0.65\gamma\langle
k_z^2\rangle$ according to our experiment of magneto-anisotropy of
electron spin dephasing.\cite{wujiang} $\dot{\rho}_{k,\sigma
\sigma^{\prime}}|_{\mbox{scatt}}$ in Eq.\ (\ref{bloch}) denote the
electron-LO-phonon, electron-AC-phonon, electron-nonmagnetic
impurity, and the electron-electron Coulomb scatterings whose
expressions are given in detail in Refs.\ \onlinecite{wu1,wu2,wu3}.
 Moreover, we further include the
spin-conserving and spin-flip electron--heavy-hole scatterings whose
expressions are given in detail in Ref.\ \onlinecite{wuzhou}. The
latter one leads to the so-called Bir-Aronov-Pikus (BAP) spin
dephasing.\cite{bap}

After numerically solving the kinetic spin Bloch equations
self-consistently, one can obtain the spin relaxation and dephasing
times from the temporal evolutions of the electron distributions and
the spin coherence.\cite{wurev}

\section{Sample growth and preparation}
Our sample was grown by molecular beam epitaxy on a
$[001]$-oriented semi-insulating GaAs substrate. The active region
is a 20 nm-wide, one-sided modulation-doped
GaAs-Al$_{0.3}$Ga$_{0.7}$As single QW. The electron density and
mobility at $T=4.2$\ K are $n_e=2.1\times 10^{11}$\ cm$^{-2}$ and
$\mu_e=1.6\times 10^6$\ cm$^2$/Vs, respectively. These values were
determined by transport measurements on an unthinned sample. For measurements in transmission geometry, the
sample was glued onto a glass substrate with an optical adhesive,
and the substrate and buffer layers were removed by selective
etching.

\section{Time-resolved Kerr and Faraday
rotation}

\subsection{Experimental setup}
For both, the TRFR and the TRKR measurements, two laser beams from a
mode-locked Ti:Sapphire laser, which is operated at 80\ MHz
repetition rate, were used. The laser pulses had a temporal length
of about 600 fs each, resulting in a spectral width of about 3-4\
meV, which allowed for a resonant excitation. The laser wavelength
was tuned to excite electrons from the valence band to states
slightly above the Fermi energy of the host electrons in the
conduction band. Both laser beams were focused to a spot of
approximately 60\ $\mu$m diameter on the sample surface. The pump
pulses were circularly polarized by an achromatic
$\frac{\lambda}{4}$ plate in order to create spin-oriented electrons
in the conduction band, with spins aligned perpendicular to the QW
plane. The TRFR measurements were performed in a split-coil magnet
cryostat with a $^3$He insert, allowing for sample temperatures
between 1.5\ K and 4.5\ K. The TRKR measurements were performed in a
continuous-flow He cold finger cryostat. In this cryostat,
non-thinned samples from the same wafer were used. Unless
otherwise stated, the experiments were carried out at a nominal
sample temperature of $T=4.5$\ K.

Average pump powers between about 100\ $\mu$W and 6\ mW were used to
create different densities, $n_{ph}$, of photoexcited, spin-aligned
electrons. The energy-dependent absorption coefficient of the sample
and the laser spot size were measured. Together with the laser beam
intensity, we estimated the total densities, $n_{ph}^{tot}$, of
electron-hole pairs, to be between about $n_{ph}^{tot}=9\times 10^9$
cm$^{-2}$ for the lowest, and $n_{ph}^{tot}=6\times 10^{11}$
cm$^{-2}$ for the highest pump intensities. Referring to $k\cdot p$
calculations of Pfalz {\em et al.},\cite{Pfalz} we have determined
for our 20 nm-wide GaAs well the densities of spin-aligned electrons
$n_{ph}$ by multiplying $n_{ph}^{tot}$ by a factor of 0.4 to account
for heavy-hole/light-hole mixing in the valence band. The resulting
maximal degree of initial spin polarization of electrons,
$P_m$, was then calculated via the relation
\begin{equation}
P_m={n_{ph}}/{(n_e+n_{ph}^{tot})}\ .
\label{poldegree}
\end{equation}
We emphasize that this value represents an upper boundary for
the initial spin polarization. In the experiment, the maximum
overlap of pump and probe beam is typically not at the beam waist,
thus we generally probe a somewhat lower density/initial spin
polarization than estimated by Eq.\ (\ref{poldegree}). In
comparing the experiment to numerical calculations, the initial spin
polarization, $P$, is thus used as a fitting parameter.
As will be shown below, we consistently found slightly lower
values of $P$, as compared to the experimentally estimated $P_m$.
The intensity of the linearly polarized probe pulses was kept
constant at an average power of about 0.5\ mW, and the rotation of
the probe polarization due to the Faraday/Kerr effect was measured
by an optical bridge.

\subsection{Absorption and power-dependent photoluminescence measurements}
In the TRFR and TRKR
measurements, various pump beam fluences are used to create
different initial values of the spin polarization. As the pump beam
fluence is increased, an increased amount of power is
 deposited in the laser focus spot, locally increasing the sample
 temperature. In order to calibrate our measurements and the
 corresponding calculations, a local probe of the sample temperature
 at the measurement spot is necessary. We utilize power-dependent
 photoluminescence (PL) measurements to determine the local sample
 temperature. For this, the sample is excited by tuning the pulsed
 Ti:Sapphire laser to slightly higher (by 14\ meV) photon energies than during
 the TRFR and TRKR measurements. In this spectral range, the absorption coefficient of
 the QW is almost constant, as determined by white-light absorption measurements.
 This enables us to observe the PL emitted from the 2DES with a grating
 spectrometer under conditions that closely resemble those during
 the time-resolved measurements.

\begin{figure}[tbh]
\begin{center}\includegraphics[width=7.6cm]{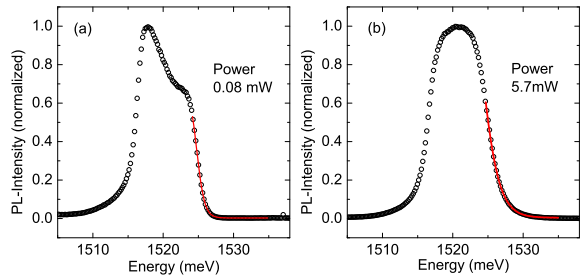}\end{center}
\caption{(color online) (Open circles) Power-dependent PL spectra
measured with a grating spectrometer. The local temperature at the
laser focus spot was determined by fitting the Fermi-Dirac
distribution function to the high-energy tail of the PL (red line).
(a) PL spectrum and fit for low pump fluence of 0.08\ mW. The
corresponding local temperature is $6.7 \pm 1$\ K. (b) PL spectrum
and fit for high pump fluence of 5.7~mW. The corresponding local
temperature is $16 \pm 2$\ K. }\label{PL_intensity}
\end{figure}

To extract the local temperature from the PL data, a Fermi-Dirac
distribution is fitted to the high-energy tail of the PL, which
corresponds to the recombination of electrons at the Fermi energy.
Figure\ \ref{PL_intensity} shows the PL data (open circles) and the
fits (red lines) for the cases of low and high pump beam fluence. In
Fig.\ \ref{PL_intensity} (a), where a low pump fluence is used, we
observe the typical, triangular shape of the PL signal from a
high-mobility 2DES, with a sharp cutoff of PL intensity for values
above the Fermi energy. In Fig.\ \ref{PL_intensity} (b),
corresponding to a large pump beam fluence, the high-energy tail is
far more rounded, indicating a higher local temperature.

\subsection{Zero-field coherent spin oscillations}

\begin{figure}[t]
\begin{center}\includegraphics[width=7.5cm]{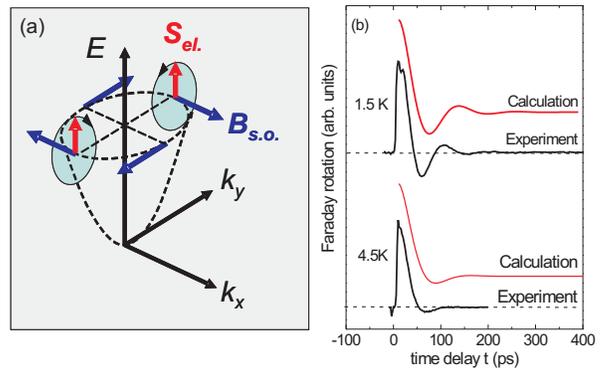}\end{center}
\caption{(color online) (a) Schematic of the geometry of the Rashba
spin-orbit field for the [001]-grown QW. For electrons at
the Fermi energy, while the direction of the Rashba field varies
within the $x$-$y$ plane, its magnitude is constant. Thus
 the $z$ component of the electron spins performs a coherent oscillation. (b)
 TRFR traces at low excitation density taken at two different temperatures. For the lower
 temperature, a coherent oscillation at \emph{zero} magnetic field is
 clearly observed. (c) Calculated spin decay curves for these two temperatures.}
 \label{Fig1}
\end{figure}

Figure \ref{Fig1} (b) shows two TRFR traces taken at \emph{zero}
external magnetic field for low pump beam fluence. The upper trace,
measured at a sample temperature of 4.5\ K, shows a strongly damped
oscillation of the TRFR signal. In the lower trace, taken at a
reduced sample temperature of about 1.5~K under otherwise identical
conditions, this damped oscillation is much more pronounced. The
oscillatory signal is due to a coherent oscillation of the excited
electron spins about an effective spin-orbit field  caused by
$k$-linear terms in the Rashba-Dresselhaus Hamiltonian. Figure\
\ref{Fig1} (a) illustrates this schematically for a pure Rashba
field: electrons are created at the Fermi energy by the pump laser
pulse, with their spins initially aligned along the growth
direction. While their $k$ vectors have arbitrary direction in the
$x$-$y$ plane, they have the same magnitude. The individual Rashba
fields for these electrons are all in-plane with the same magnitude,
causing the electron spins to precess into the sample plane with
equal Larmor frequencies. The observed oscillation of the TRFR
signal is the coherent sum of the $z$ component of the individual
spins oscillating about their individual Rashba-Dresselhaus fields.
This coherent oscillation has already been observed
experimentally\cite{Brand} and is representative of the weak
scattering limit. We note that the amplitude of the effective
Rashba-Dresselhaus field, calculated from the oscillation frequency
using the in-plane electron g factor $|g|= 0.355$, is $B_{eff} =
2.35$\ T. The red lines in Fig.\ \ref{Fig1} (b) present the
calculated temporal evolutions of the differences of spin -up and
-down electron densities (normalized $\Delta N$) for the two
corresponding cases. One can see that the damped oscillation is
indeed pronounced for lower temperature. Furthermore, the
oscillation period is very sensitive to the strength of
Rashba/Dresshause spin-orbit coupling and electron momentum
scattering time. Therefore, it is understandable that the calculated
oscillation period is a little different from the experiment as all
the parameters we used are fixed.

\subsection{Dependence of SDT  on initial spin
polarization}

\begin{figure}[tbh]
\begin{center}\includegraphics[width=8.6cm]{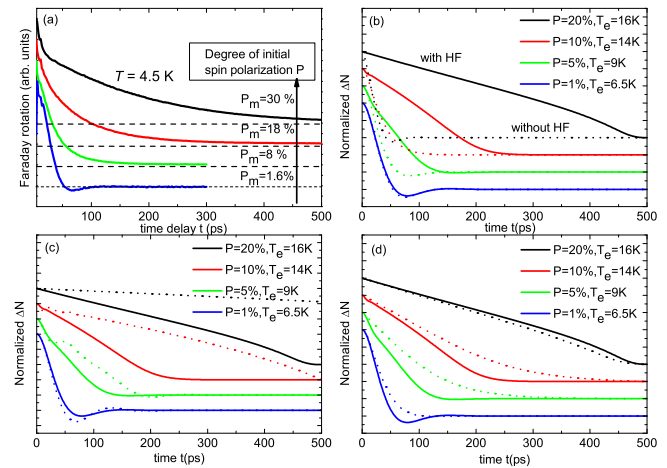}
\end{center}
\caption{(color online) (a) TRFR traces for different pump beam
fluences and therefore different initial spin polarizations. $P_m$
is the maximum initial spin polarization calculated from Eq.\
(\ref{poldegree}). (b) Calculated spin polarization decay curves for
different initial spin polarizations $P$ with (solid curve) and
without (dashed curve) HF term. The free parameters in the
calculations are the initial spin polarization, and we present the
optimal fitting parameters. (c) Calculated spin polarization decay
curves with (solid curve) and without (dashed curve)  the
electron-hole Coulomb scattering. (d) Calculated spin polarization
decay curves with (solid curve) and without (dashed curve)  the
screening from holes. }\label{Leistung}
\end{figure}

Figure \ref{Leistung} (a) shows a series of TRFR traces taken for
different pump beam fluences and thus different initial values of
the spin polarization. For all TRFR traces, a very fast decay of the
TRFR signal is observed during the first few picoseconds after
excitation. We attribute this to the spin polarization of the
photoexcited holes, which typically lose their initial spin
orientation extremely fast. A second, significantly slower decaying
part of the signal is attributed to the spin dephasing of the
photoexcited electrons. Using a biexponential fit function, the SDT
is determind from the data.  It is clearly visible that with
increasing spin polarization, the SDT increases as well, from about
20~ps to more than 200~ps. This observation is in good agreement
with predictions by Weng and Wu,\cite{wu1} which stem from their
fully microscopic calculations. It has been mentioned above that the
estimated value of initial spin polarization, $P_m$, in Eq.\
(\ref{poldegree}) is an upper boundary. Therefore, the polarization
values we used in Ref.\ \onlinecite{stich}
[cf.\ Fig.\ \ref{Leistung} (a)] are actually
larger than the real ones. For this reason, we introduced in
Ref.\ \onlinecite{stich} a fitting parameter $\tau$ to obtain the
same $T_{2}^{\ast}$ as the experiment. In this paper, we choose the
initial spin polarization $P$ as a fitting parameter instead of
introducing $\tau$. This seems to be more reasonable, since, as
mentioned above, the experimentally determined $P_m$ is just an
upper boundary.  Moreover, the hot electron temperatures,
$T_e$, are obtained from PL spectra to be 6.5\ K, 9\ K, 14\ K, and
16\ K for the experimental traces, displayed in Fig.\ \ref{Leistung}
(a). These values are used in the calculations. In Fig.\
\ref{Leistung} (b), the temporal evolutions of the spin polarization
resulting from calculations with (solid lines) and without the
HF term (dashed lines) are compared to the experimental results,
showing an excellent agreement with the best fitting parameters. It
is noted that the same parameters are used for both calculations
with and without the HF term. Obviously, the increase of SDT with
increasing $P$ originates from the HF term.

Moreover, we present the temporal evolution of spin polarization
with (solid lines) and without (dashed lines)
spin-conserving electron--heavy-hole Coulomb scattering in Fig.\
\ref{Leistung} (c). (The detail of electron-heavy hole scattering
terms of can be found in Ref.\ \onlinecite{wuzhou}.) It is obvious
that scattering strength is enhanced while including the
spin-conserving electron--heavy-hole scattering. Also the larger the
hole density (which increases with the pumped spin polarization) is,
the larger the enhancement of scattering strength is. Therefore, the
SDT is reduced by the spin-conserving electron--heavy-hole Coulomb
scattering. This is
 consistent with the effect of the scattering in the weak scattering limit.\cite{lvchengwu}
 In addition, we have also investigated
the effect of the spin-flip electron--heavy-hole Coulomb scattering
(the BAP mechanism). We do not present the corresponding figure in
this manuscript due to the fact that the BAP mechanism hardly
changes the temporal evolution of the spin polarization and can be
ignored in our cases as studied by Zhou and Wu recently.\cite{wuzhou}

Furthermore, in Fig.\ \ref{Leistung} (d), we show the temporal
evolution of spin polarization with (solid lines) and without
(dashed lines) the screening from the holes in the screened
Coulomb potential under the random-phase approximation (detailed
expression can also be found in Ref.\ \onlinecite{wuzhou}). There
are two mechanisms from the hole screening that influence  the spin
dephasing. On one hand, the presence of hole screening strengthens
the total screening and therefore reduces the electron-electron and
electron-hole Coulomb scattering. This leads to an increase of the
SDT. On the other hand, the presence of hole screening reduces the
effect of the HF term. This leads to a reduction of the SDT. The
competition between these two mechanisms is clearly shown in the
figure: For the lowest spin polarization, the HF term is not
important. Therefore the first mechanism is dominant; For the other
three higher polarizations, the HF term is large enough, which leads
to the domination of the second mechanism.

In order to verify that the increased SDTs observed
in the measurements shown above are due to the increase of the initial
spin polarization, instead of either due to the increased electron
density or due to the sample heating as the pump beam fluence is increased,
measurements with constant excitation density were performed. To
vary the degree of initial spin polarization independently of
excitation density, the circular polarization degree of the pump
beam was adjusted by rotating the $\frac{\lambda}{4}$ plate in the
pump beam. The circular polarization degree as a function of the
$\frac{\lambda}{4}$ plate angle was measured by using a second
$\frac{\lambda}{4}$ plate and a polarizer to analyze the pump beam
polarization state.
\begin{figure}[thb]
\begin{center}\includegraphics[width=7.5cm]{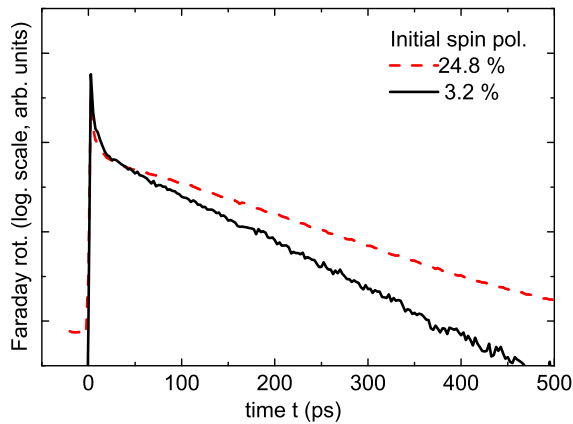}\end{center}
\caption{(color online)  Two TRFR traces for different initial spin
polarizations, which were created by varying the circular
polarization degree of the pump beam.}\label{Fig4}
\end{figure}

\begin{figure}[thb]
\begin{center}\includegraphics[width=7.5cm]{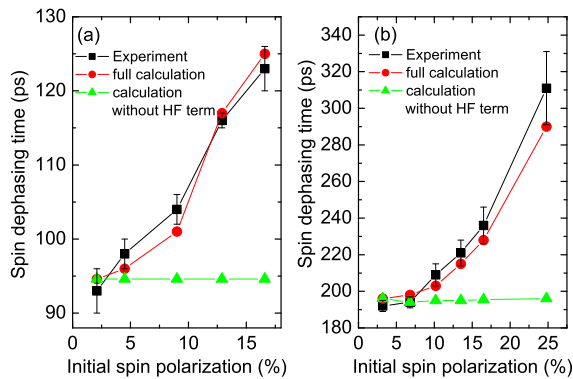}\end{center}
\caption{(color online) (a) The SDTs as a function of initial spin
polarization for constant, \emph{low} excitation density and
variable polarization degree of the pump beam. The measured spin
dephasing times are compared to calculations with and without the HF
term, showing its importance. (b) The SDTs measured and calculated
for constant, \emph{high} excitation density and variable
polarization degree. The values for lowest and highest initial spin
polarization correspond to the TRFR traces shown in figure
\ref{Fig4}.}\label{SpinDeg}
\end{figure}

Figure\ \ref{Fig4} shows two TRFR traces for low initial
spin polarization generated by a nearly linearly-polarized pump
beam, and high initial spin polarization generated by a
circularly-polarized pump beam, using the same, high pump beam
fluence, and thus resulting in identical electron density and
temperature. The traces were normalized to allow for easy
comparison. It is clearly visible that the spin dephasing time is
longer for the high initial spin polarization case. In Fig.\
\ref{SpinDeg} (b) the SDTs for a series of
measurements with constant, high pump beam fluence and varying
initial spin polarization degree are shown, clearly demonstrating an
increase of the SDT from less than 200\ ps for the
low-initial-polarization case to about 300\ ps for high initial
polarization. They are compared to calculation with and without the
HF term. The calculations including the HF term are in excellent
agreement with the measured data for both low and high excitation,
which again show an increase of the SDT with rising initial spin
polarization. If the HF term is excluded from the calculations, the
spin dephasing term is nearly independent of initial spin
polarization.

Figure\ \ref{SpinDeg} (a) compares the SDTs of a second
series of measurements with constant excitation
densities and variable initial spin polarization to calculations
with and without the HF term. In this measurement series, low pump
beam fluence,  and hence low total carrier density, was used.
In the calculation, except for the largest initial spin polarization
in each case, there is no fitting parameter. Again, excellent
agreements are obtained between the experiment and the theory.
Moreover, our results show that the increase of the SDT does solely
stem from the HF contribution instead of other effects.

\subsection{Dependence of $g$ factor on initial spin polarization}

\begin{figure}[tbh]
\begin{center}\includegraphics[width=7.5cm]{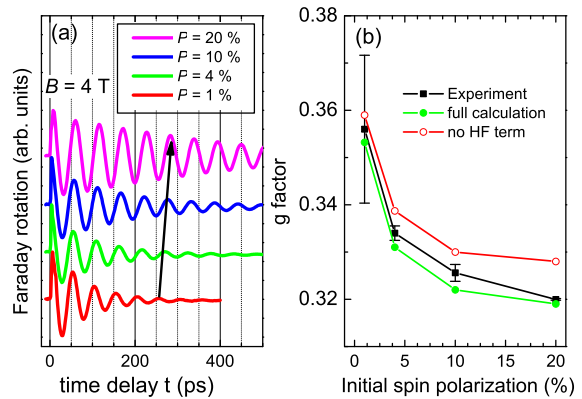}\end{center}
\caption{(color online) (a) TRFR measurements at $B=4$\ T for
different $P$. An increase of the electron precession period with
increasing $P$ is clearly observed (arrow). (b) Comparison of
electron $g$ factors for different polarization degrees $P$, as
extracted from the experiments (solid squares), and the calculations
with (solid dots) and without (open dots) HF term.}\label{gFactor}
\end{figure}

To determine the electron $g$ factor of the 2D electron system, TRFR
measurements with a magnetic field applied within the sample plane
were performed (Voigt geometry). The $g$ factor was extracted from
the precession frequency as a function of the applied magnetic
field. Figure\ \ref{gFactor} (a) shows TRFR traces, taken with a
magnetic field of 4\ T applied in the sample plane. The pump laser
fluence was varied, resulting in different initial spin
polarizations and electron densities. With rising initial spin
polarization, the effective $g$ factor is reduced by about 10
percent. In Fig.\ \ref{gFactor} (b), the experimental results are
compared to the calculations with and without the HF term, where the
same values for the spin polarization  as in Fig.\ \ref{Leistung}
were used. The calculations show a similar decrease of the $g$
factor with increasing spin polarization, and the HF term provides
only a small correction.

\subsection{Temperature dependence for different initial spin polarizations}

\begin{figure}[thb]
\begin{center}\includegraphics[width=7.5cm]{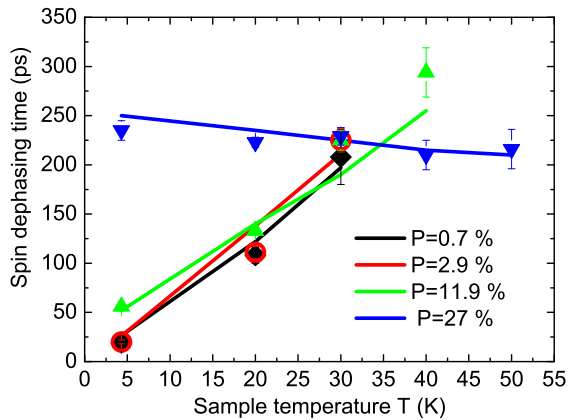}\end{center}
\caption{(color online)
 The Spin dephasing time as a function of sample temperature, for
 different initial spin polarizations. The measured data points are represented by
 solid points, while the calculated data are represented
 by  lines of the same colour.}\label{Fig7}
\end{figure}

\begin{figure}[thb]
\begin{center}\includegraphics[width=7.5cm]{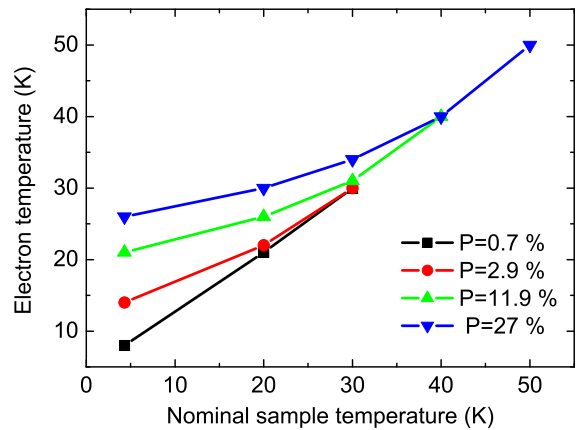}\end{center}
\caption{(color online)
 The electron temperature determined from intensity-dependent PL measurements as
 a function of the nominal sample temperature, for different pump beam fluence
 and initial spin polarization, under experimental conditions corresponding to
 the measurements shown in figure \ref{Fig7}. The measured data points are represented by
 solid points, while the lines serve as guide to the eye.}\label{Fig8}
\end{figure}

Temperature-dependent measurements were performed in a He-flow
cryostat in reflection (Kerr) geometry. The SDT, which were
determined by fitting a biexponential decay function to the
experimental data, are shown in Fig.\ \ref{Fig7} as a function of
temperature for different pump beam intensities, i.e., different
spin polarizations $P$.  The theoretical calculations (solid lines
of the same color) are in good agreement with the experimental
results. We stress that the only fit parameter for each series is
the value of the spin polarization, $P$, which was adjusted to
reproduce the experimental data point at highest temperature. Then,
the respective temperature dependencies were calculated, keeping $P$
and all other parameters fixed.  The hot electron temperatures are
taken from the experimental values determined by the
intensity-dependent PL measurements. For high pump beam fluence, the
electron temperature is significantly higher than the nominal sample
temperature, especially for the lower sample temperatures, as Fig.\
\ref{Fig8} shows. It is clearly visible in Fig.\ \ref{Fig7} that for
the small initial spin polarization, the SDT drastically increases
as the sample temperature is raised, for instance, from about 20\ ps
at 4\ K to 200\ ps at 50\ K for $P=0.7$\ \%. Remarkably, for large
initial spin polarization, the SDT {\em decreases} with temperature
from about 250\ ps at 4\ K to a little more than 210\ ps at 50\ K
for $P=16$\ \%. These features again agree with the theoretical
predictions.\cite{wu1} For small spin polarization, a large increase
of the SDT with rising temperature has already been observed by
Brand {\em et al.}.\cite{Brand} This behavior has been discussed
from kinetic spin Bloch approach by Weng and Wu\cite{wu1,wu4} in the
high temperature regime and by Zhou {\em et al}.\cite{wu3} in the
low-temperature regime. It is due to the increase of the momentum
scattering with temperature that leads to the increase of the SDT in
the strong scattering limit.\cite{lvchengwu} For large spin
polarization, the decrease of SDT is due to the fact that the
effective magnetic field from the HF term decreases with
temperature.

\section{Conclusion}
In conclusion, we have performed time-resolved Kerr and Faraday
rotation measurements on a high-mobility two-dimensional electron
system at low temperatures. We observe that the SDT strongly depends
on the initial spin polarization within the sample. This effect is
due to the HF term of the electron-electron Coulomb interaction,
which serves as an effective magnetic field along the growth axis
and inhibits the spin dephasing. By independently varying the degree
of initial spin polarization while keeping the excitation density in
the experiment constant, we can clearly exclude unrelated origins of
the observed increase in spin dephasing time. Furthermore, the
contributions of the spin-conserving, spin-flip electron--heavy-hole
scattering, and the Coulomb screening from the photo excited hole
gas to spin dephasing are studied. The spin-conserving
electron--heavy-hole scattering makes the SDT shorter; the spin-flip
process can be ignored; and the hole screening makes the SDT larger
for small spin polarization and smaller for large ones. Moreover,
the electron $g$ factor decreases with increasing spin polarization
which is both observed experimentally and reproduced theoretically
in the calculations.  Finally, we find that the temperature
dependence of SDT are very different for small and large spin
polarizations. For small spin polarization, the SDT increases with
temperature; and for large one, it decreases. Both are in good agreement
with the theoretical predictions. In the theory, except for the
large initial spin polarization which can not be fully determined
from experiment and is treated as fitting parameter, all the other
parameters are taken from the experiments. The calculated results fit pretty well with
the experimental data. This indicates that the approach based
on the kinetic spin Bloch equation can be used in calculating the
spin dynamics quantitatively.

\section{Acknowledgements}
We gratefully thank Jaroslav Fabian and R. T. Harley for valuable
discussions. This work was supported by the Deutsche
Forschungsgemeinschaft via GrK 638, grant No. Schu1171/1-3,
Schu1171/5-1, and SFB 689, the Natural Science Foundation of China
under Grant No.\ 10574120, the National Basic Research Program of
China under Grant No.\ 2006CB922005, the Knowledge Innovation
Project of Chinese Academy of Sciences and SRFDP. One of the authors
(M.W.W.) would like to thank Hailin Wang in University of Oregon,
USA, for hospitality where this work was finalized.

\end{document}